\documentclass[twocolumn,aps,prl,twoside,showkeys]{revtex4}

\usepackage{graphicx}        
\usepackage{amsbsy}

\usepackage{mathptmx,courier,helvet}

\usepackage[colorlinks=true,citecolor=blue]{hyperref}

\begin{document}

\preprint{ICPS Vienna 2006}

\title{A solution in 3 dimensions for current in a semiconductor
   under high level injection from a point contact}
\author{Mike W. Denhoff}
\email{mike.denhoff@nrc.ca}
\affiliation{National Research Council, Institute for 
Microstructural Sciences, M-50, Montreal Road, Ottawa, ON, K1A~0R6, Canada}
\date[Published in: ]{\textit{28th Intl. Conf. on the Physics of Semiconductors},
  Vienna, 24-28 July 2006, AIP Conference Proceedings Vol. 893.}


\begin{abstract}
The standard equations for semiconductor device analysis were solved
by specifying the electron and hole current injected at a
small contact, assuming high-level injection.
Calculated current-voltage characteristics were fit to
measurements of a single point breakdown in an ultrathin dielectric.
It was found that the minority carrier injection level was about
70\%.
\end{abstract}
\maketitle

\section{Introduction}
The nature of the contact to a semiconductor at
point breakdowns in ultrathin oxides, is not well
understood.  An important observed feature of these contacts is that
they can have a high minority carrier injection ratio.  There have been
a number of attempts to model point breakdowns in ultrathin oxides
on silicon~\cite{miranda}.  These models concentrate on the physics at the contact
interface and attempt to derive the current vs voltage ($IV$) characteristic.
However, in many physical devices the main contribution to the $IV$
characteristic is due to the conduction in the bulk of the semiconductor.
Therefore, one should include a model of the bulk conductivity in an
explanation of point contact behavior.  In this paper, such a model is
developed using standard semiconductor equations solved for
point contacts with minority carrier injection.

\section{Point contact physics}
This problem has the same geometry as the spreading resistance problem.
The difference is that the spreading resistance problem assumes an
ohmic contact and only majority carrier conduction.  In the point contact
problem arbitrary levels of minority carrier injection will be included.
This calculation will assume high-level injection,
which implies that there is charge neutrality in the bulk. 
This means that the number of excess holes (beyond the 
equilibrium concentration) is equal to the number of excess electrons.
Since the bulk is neutral, the Laplace equation can be used to
find the electric field,
\begin{equation}  
\nabla\cdot\mathbf{E} = 0.
\end{equation}
The remaining equations needed to describe conduction in a semiconductor
are the continuity equations for holes and electrons along with expressions
for the hole and electron current in terms of drift and 
diffusion~\cite{nussbaum,sze}.  For high-level injection, as in the case
of high forward bias of a pn junction~\cite{nussbaum}, these equations 
reduce to
\begin{equation}
\nabla^2 p_t - \frac{p}{L_p^2} = 0,\quad \nabla^2 n_t - \frac{n}{L_n^2} = 0,
\label{eq:pcont}
\end{equation}
for holes and electrons.
$p_t$ ($n_t$) is the total hole (electron) concentration, $p$ ($n$) is
the excess hole (electron) concentration, and $p_0$ ($n_0$) is the
equilibrium hole (electron) concentration, so that
$p_t=p+p_0$ ($n_t=n+n_0$).
$L_p$ is the high-level diffusion length,
$L_p^2 = 2\mu_n D_p \tau_p/(\mu_n+\mu_p)$,
where $\mu_p$ and $\mu_n$ are hole and electron mobilities, $D_p$
is the hole diffusion constant, and $\tau_p$ is the hole lifetime.
A similar equation can be written for $L_n$.
Since $\mu_pD_n = \mu_nD_p$ and $\tau_p = \tau_n$, then
$L_p = L_n = L$.

These equations must be solved for the boundary conditions of the contact.
A semi-infinite structure will be used, where the surface of the semiconductor
is on the $x$-$y$ plane and the bulk of the semiconductor extends in the
positive $z$ direction to infinity.  The potential at infinity will be taken
as 0.  There are mixed boundary conditions on the $x$-$y$ plane.
Similar mixed boundary conditions, for the Laplace equation, are
used in the solution of the usual spreading resistance
problem~\cite{tranter,denhoff}.
Outside the contact radius $a$, 
the current through the surface and the perpendicular component of the
electric field are 0.
The boundary conditions on the contact are given by specifying the 
hole and electron currents through the contact.
These conditions are set on the semiconductor side of the contact
in order to avoid the physics of the contact interface.

\section{3-d solution}

The problem is cylindrically symmetric about the $z$-axis which is 
perpendicular to the center of the contact.  Equation~(\ref{eq:pcont})
can be written (ignoring the $\theta$ dependence) in cylindrical
coordinates,
\begin{equation}
\frac{\partial^2p}{\partial r^2}
   + \frac{1}{r} \frac{\partial p}{\partial r}
     + \frac{\partial^2p}{\partial z^2}
      - \frac{p}{L^2} = 0.
\end{equation}
This is the Helmholtz equation.  It can be solved by separating
variables.  Using the boundary condition that the potential is zero
for $r$, $z$ at infinity, the solution for excess hole concentration is
\begin{equation}
p(r,z) = \int_0^\infty A(\lambda) J_0(\lambda r)
        \exp{(-[\lambda^2+1/L^2]^{\frac{1}{2}}\,z)} \mathrm{d}\lambda.
\end{equation}
$J_0$ is a Bessel function of order 0.
The function $A(\lambda)$ must be found using the remaining boundary
conditions.
A similar solution for the electron density can be written.

An analytic solution can be found if $A(\lambda)$ is chosen to be
\begin{equation}
A(\lambda) = B\frac{J_1(a \sqrt{\lambda^2+c^2})}{(\lambda^2+c^2)}\lambda,
\end{equation}
where $J_1$ is a Bessel function of order 1, $B$ is a constant, 
and $c^2 = 1/L^2$.
With this choice, the current density over the area of the contact
is approximately constant for $a\ll L$. 
The electric field at the contact must be consistent with the current,
which implies that the on the area of the contact the $z$ component of
the field is
\begin{equation}
E_z = EJ_0(c\sqrt{a^2+r^2}), \quad \mathrm{for}\ r<a
\end{equation}
where $E$ is a constant.
Since,
in the high-level injection case, $p=n$, the solutions can be equated
and values of $B$ and $E$ determined for any given values of total
hole current, $I_h$, and electron current, $I_n$.  The potential
at the center of the contact can be found from the solution of the Laplace 
equation and thus the $I$ vs $V$ curve can be calculated.   
The numerical results below use typical values of parameters for silicon.

\subsection{Injection of electrons}

An interesting case is that of injecting a current of
pure electrons into a p-type silicon.  
In the case of large injected current,
\begin{figure}
{\includegraphics[width=8.5cm]{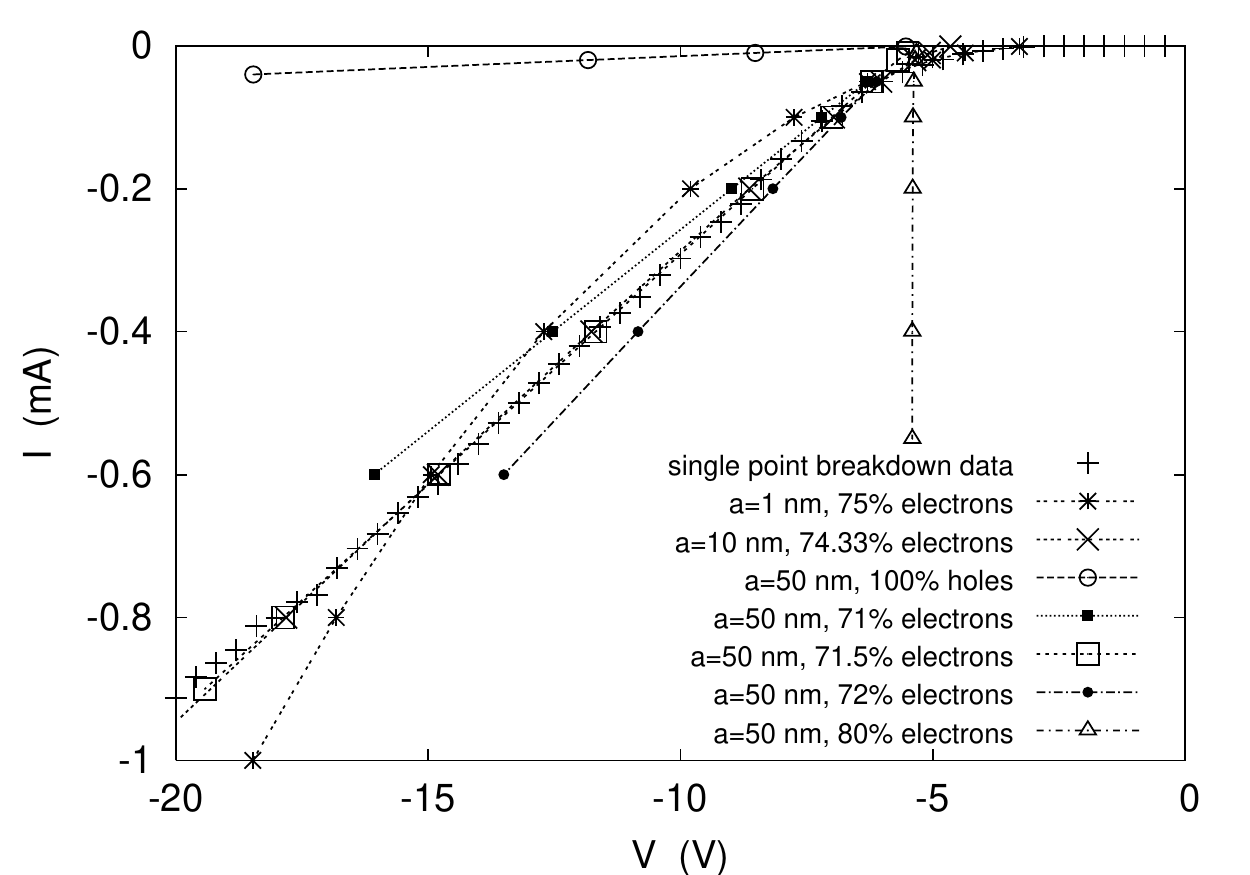}}
\caption{\label{iv-a5b}$IV$ data of a single point breakdown.  
Theory curves have voltage offsets of -2.8\,V, -4.5\,V, and -5.2\,V for 
$a=1$\,nm, $a=10$\,nm, and $a=50$\,nm, respectively.}
\end{figure}
the solution for $E$ and the resistance can be approximated 
as
\begin{equation}
E = \frac{-D_n}{2aG\mu_n},\qquad R = \frac{0.85 D_n}{2 G\mu_nI_n}.
\end{equation}
$G$ is somewhat dependent on $a$, but is approximately
equal to $4/(3\pi)$.
It is seen that as the electron current gets large, the
electric field coefficient tends to a constant.  This is
opposed to the normal Ohm's law behavior where the electric
field is proportional to current.  The resistance of the contact
actually decreases as $1/I_n$ as opposed to the ohmic behavior
of a constant resistance.

\subsection{Single point breakdown}
A single point breakdown device was fabricated by opening a 100\,nm
square window in a thick (20\,nm) SiO$_2$ insulating layer on a
6\,$\Omega\cdot$cm p-type silicon wafer.  A thin SiO$_2$ layer (2\,nm) was grown
in this opening and an indium tin oxide top electrode was deposited on top
of this~\cite{mihaychuk}.  The device was voltage stressed to
create a hard breakdown.
IV data of this breakdown site is presented in
Figure~\ref{iv-a5b}.
Theoretical $IV$ curves
 were calculated assuming a constant
electron injection ratio.
A constant voltage was added to model
the voltage drop across the contact interface for large current.
The experimental current is more than 10 times larger than the calculated
current for only hole injection.
Considering the curvature at low currents, the best fit is for $a=10$\,nm
and an electron injection of 74\%.
With smaller values of $a$ the theoretical curve is not straight
and cannot fit the measured data.
A reasonable fit can be obtained for a range of contact size, however it is
clear that the electron injection ratio is between 70\% and 75\%.
This demonstrates that a large minority injection ratio exists for these
devices.

\end{document}